\documentclass[seceq,preprint]{ptptex}

\notypesetlogo                       
\preprintnumber[3cm]{
KEK-TH-1130\\ December, 2006}

\title{
Electric Dipole Moment of Magnetic Monopole%
}

\author{
Makoto \textsc{Kobayashi}%
}

\inst{
High Energy Accelerator Research Organization (KEK), \\Tsukuba 305-0801, Japan \\ and \\
        International Institute for Advanced Studies,\\Kizu 619-0225, Japan   
}

\abst{
The electric dipole moment of a magnetic monopole with spin is studied in the $N=2$ supersymmetric
gauge theory. Dipole moments of electric charge distributions, as well as dipole moments
due to magnetic currents, are calculated. The contribution of the charge distribution of the 
fermion to the gyroelectric ratio is expressed in terms of $\zeta(3) $.
}

\begin{document}

\maketitle

\section{Introduction}

Magnetic monopoles appear as soliton solutions in certain kinds of gauge theories.
According to the duality idea, these monopoles are expected to behave as fundamental particles
in a dual description of the system.
Therefore, it is interesting to investigate the extent to which magnetic monopoles share the properties 
of electrically charged particles. In connection to this point, the electric dipole moment of monopoles
has been attracting attention\cite{Montonen, Osborn, Kastor}.

If a magnetic monopole has spin, it will have an electric dipole moment proportional to its spin angular momentum,
\begin{equation}
\vec{d}=-g_{m}\frac{Q_m}{2M}\vec{J},
\end{equation}
where $g_m$ corresponds to the $g$-factor of the magnetic moment of electrically charged particles.
We conjecture that $g_m$ is 2 for the spin 1/2 monopole, with possible radiative corrections.
Actually, it is known that $g_m$ is 2 for BPS monopoles of the $N=2$ supersymmetric 
gauge theory.\cite{Osborn, Kastor}
This can be shown by considering the long-distance behavior of the electric field of a spin 1/2 monopole,
which is obtained by applying a finite supersymmetry transformation to the spinless monopole solution.
However, the properties of the electric dipole moments of monopoles are yet to be explored in detail.
In this paper,  the source term of the electric field is investigated to clarify how the dipole field is generated.
We find, for example, that the electric dipole moment of the charge distribution of the fermion field
is given by a curious number and also that 1/3 of the dipole moment is generated by magnetic currents.

\section{$N=2$ supersymmetric gauge theory}

In this section, we summarize the basic properties of magnetic monopoles in the $N=2$ supersymmetric gauge theory.\cite{Harvey}
These properties are useful for the calculation given in the following sections.The gauge group is assumed to be $SU(2)$, 
and all the fields belong to the adjoint representation.

The Lagrangian is given by
\begin{equation}
\begin{split}
\mathcal{L}=&\displaystyle{{\rm Tr}\Bigl(-\frac{1}{4}F_{\mu\nu}F^{\mu\nu}
            +\frac{1}{2}D^{\mu}SD_{\mu}S+\frac{1}{2}D^{\mu}PD_{\mu}P
            +\frac{e^2}{2}[S,P]^2}\\
           &\displaystyle{+i\bar{\psi}\gamma^{\mu}D_{\mu}\psi
           -e\bar{\psi}[S,\psi]+ie\bar{\psi}\gamma_5[P,\psi]\Bigr)} ,\label{eq:lagrangian}
\end{split}
\end{equation}
where the following matrix notation is adopted,
\begin{equation}
F_{\mu\nu}=F^a_{\mu\nu}t^a,\hspace{0.5cm}S=S^at^a,\hspace{0.5cm}
P=P^at^a,\hspace{0.5cm}\psi=\psi^at^a,
\end{equation}
with
\begin{equation}
(t^a)_{bc}=\epsilon^{abc}.
\end{equation}
The field strength and the covariant derivative are defined as
\begin{equation}
F_{\mu\nu}=\partial_{\mu}A_{\nu}-\partial_{\nu}A_{\mu}
             -ie[A_{\mu},A_{\nu}],
\end{equation}
\begin{equation}
D_{\mu}\ast =\partial_{\mu}\ast -ie[A_{\mu},\ast ]. 
\end{equation}
We also use the following notation:
\begin{equation}
E^i=F_{0i},\hspace{1cm}B^i=-\frac{1}{2}\epsilon^{ijk}F_{jk}. 
\end{equation}
The Lagrangian (\ref{eq:lagrangian}) is invariant, up to a total derivative term, under the
$N=2$ supersymmetry transformations
\begin{align}
\delta A_{\mu} &=i\bar{\alpha}\gamma_{\mu}\psi
                 -i\bar{\psi}\gamma_{\mu}\alpha, \\
\delta S &=i\bar{\alpha}\psi
                 -i\bar{\psi}\alpha, \\   
\delta P &=\bar{\alpha}\gamma_5\psi
                 -\bar{\psi}\gamma_5\alpha, \\ 
\delta \psi &=(\frac{1}{2}\sigma^{\mu\nu}F_{\mu\nu}-\gamma^\mu  D_\mu S
                +i\gamma^\mu  D_\mu P\gamma_5-e[P,S]\gamma_5)\alpha,\label{eq:delpsi}
\end{align}
where the parameter $\alpha $ is a Grassmann-valued Dirac spinor.
The equations of motion for this system are as follows:
\begin{align}
D_{\mu}F^{\mu \nu } -ie[P,D^{\nu }P]-ie[S,D^{\nu }S]-e[\bar{\psi },\gamma^{\nu }\psi   ]&=0,\label{eq:field}\\
D_\mu D^\mu S-e^2[P,[S,P]]-e[\bar{\psi },\psi  ]&=0,\\
D_\mu D^\mu P-e^2[S,[P,S]]+ie[\bar{\psi }\gamma_5 ,\psi  ]&=0,\\
i\gamma^\mu D_\mu \psi -e[S,\psi ]+ie\gamma_5[P,\psi ]&=0.\label{eq:dirac} 
\end{align}

It is known that this system has a BPS monopole, which is obtained by solving the BPS equation,
\begin{equation}
-\frac{1}{2}\epsilon^{ijk}F_{jk} =D_iS,
\end{equation}
together with@
\begin{equation}
A_0=P=\psi =0.\label{eq:A0andP}
\end{equation}
The explicit form of the solution is given by
\begin{align}
S^a&=\hat{r}^a\frac{H(evr)}{er},\label{eq:higgs}\\ 
A^a_i&=\epsilon^{aij}\hat{r}^j\frac{1-K(evr)}{er}\label{eq:gauge}, 
\end{align}
where
\begin{align}
H(y)&=y\ {\rm coth}\; y-1, \label{eq:H}\\
K(y)&=\displaystyle{\frac{y}{{\rm sinh}\; y}}  \label{eq:K}.
\end{align}

The above solution corresponds to a spinless monopole.
A spinning monopole is obtained by creating a zero energy mode fermion in the spinless monopole background.
Owing to supersymmetry, the zero energy mode can be obtained by applying supersymmetry transformations to 
the spinless monopole configuration.

Substituting Eqs. (\ref{eq:A0andP}), (\ref{eq:higgs}) and (\ref{eq:gauge}) into Eq. (\ref{eq:delpsi}), we have
the following variation of the fermion field in the monopole background: 
\begin{equation}
\delta \psi =\Bigl(\frac{1}{2}\sigma^{ij}F_{ij}-\gamma^iD_iS\Bigr)\alpha =-2\gamma^i D_iSP_+\alpha. \label{eq:delpsi0}
\end{equation}
Here we have used the representation 
\begin{equation}
\gamma^0=\biggl(\begin{array}{cc} 0 &-i \\
                                  i & 0  \end{array}\biggr), \hspace{1cm}
\gamma^k=\biggl(\begin{array}{cc} -i\sigma^k  &   0 \\
                                     0        & i\sigma^k   \end{array}\biggr), 
\end{equation}
for the gamma matrices and $P_\pm$ is defined as
\begin{equation}
P_\pm =\frac{1}{2}(1\pm \Gamma^5), 
\end{equation}
with
\begin{equation}
\Gamma^5=-i\gamma^0\gamma^5=\biggl(\begin{array}{cc} 1 &  0 \\
                                                     0 & -1  \end{array}\biggr) .
\end{equation}
We note that $\delta \psi $ in Eq. (\ref{eq:delpsi0}) is a zero energy solution to Eq. (\ref{eq:dirac}) 
in the background field of the spinless monopole configuration.
It is known that the back reaction of the created fermion can be described by considering the finite
supersymmetry transformation given by\cite{Aichelburg, Kastor}
\begin{align}
\tilde{\psi}&= -2\gamma^k D_kSP_+\alpha,  \label{eq:psitilde}\\
\tilde{A_0}&=\tilde{P}\ =\ -2iD_kS(\alpha^\dagger \gamma^kP_+\alpha  ), \label{eq:a0tilde}\\
\tilde{A_i}&=A_i, \label{eq:aitilde}\\
\tilde{S}&=S,  \label{eq:stilde}
\end{align}
where the transformed fields are indicated by tildes.
We can directly confirm that they satisfy the equations of motion exactly.

It is convenient to interpret $P_+\alpha $ as the operator, instead of the supersymmetry transformation parameter,
which satisfies the anti-commutation relation\cite{Kastor}
\begin{equation}
\{\chi^a, \chi^\dagger_b \}=\frac{1}{4M}\delta^a_b,    
\end{equation}
where $\chi$ is defined through
\begin{equation}
\alpha =\biggl(\begin{array}{c}\chi \\ 0 \end{array}\biggr).
\end{equation}
The spin 1/2 monopole states are obtained by applying the creation operators $\chi^\dagger$
to the spinless monopole state.
We note that the spin angular momentum of the monopole can be expressed in this formalism as 
\begin{equation}
J^k=2iM(\alpha^\dagger \gamma^k\alpha   )=2M(\chi^\dagger \sigma^k\chi ) . 
\end{equation}

Now we can calculate the electric dipole moment of the spin 1/2 monopole in the following way.
From Eqs. (\ref{eq:a0tilde}) and (\ref{eq:aitilde}), 
we find that the $i0$-component of the guage field configuration is expressed as
\begin{equation}
\tilde{E}^i=\tilde{F}^{i0}=-D_i\tilde{A}_0=2iD_iD_kS (\alpha^\dagger \gamma^kP_+\alpha  ).
\end{equation}
The electric field is obtained by projecting $\tilde{E}^{ai}$ onto the direction of $\tilde{S}^a$.
After some calculations, we find that the electric field exhibits the following dipole behavior in the asymptotic region:
\begin{equation}
\hat{r}^a\tilde{E}^{ai}\rightarrow -\frac{1}{eMr^3}(3\hat{r}^i\hat{r}^k-\delta^{ik})J^k .    
\end{equation}
From this, we conclude that the $g$-factor of the spin 1/2 monopole is given by\cite{Kastor}
\begin{equation}
g_m=2.
\end{equation}
In the following sections, we investigate how this dipole field is generated.

\section{Calculation of the electric dipole moment}

Our primary interest is to determine the dipole moment of an electric charge distribution.
For this purpose, we first define the charge density.
Projecting the time component of Eq. (\ref{eq:field})
onto the direction of the scalar field, we have
\begin{equation}
\partial_i(\hat{S}^aE^{ai} )=\rho_{A1}+\rho_{A2} +\rho_S +\rho_P+ \rho_\psi \label{eq:Gauss},   
\end{equation}
where
\begin{align}
\rho_{A1} &= (\partial_i \hat{S}^a )E^{ai} , \\
\rho_{A2} &= -e\epsilon^{abc}\hat{S}^aA_i^bE^{ci} ,   \\  
\rho_S &= -e\epsilon^{abc}\hat{S}^aS^b(D^0S)^c ,  \\
\rho_P &= -e\epsilon^{abc}\hat{S}^aP^b(D^0P)^c ,  \\
\rho_\psi  &= ie\epsilon^{abc}\hat{S}^a\bar{\psi }^b\gamma^0\psi ^c,   
\end{align}
and
\begin{equation}
\hat{S}^a=\frac{S^a}{|S|}.
\end{equation}
It is natural to define the right-hand side of Eq. (\ref{eq:Gauss}) as the charge density. 
It consists of terms which express the contribution of each field. 
We note that $\rho_{A1}+\rho_{A2}$, $\rho_{S}$, $\rho_{P}$ and $\rho_{\psi}$ are 
gauge invariant.
The quantity $\rho_{A1} $ arises from the exchange of the derivative and $\hat{S}^a$.

Substituting the field configuration for the spin 1/2 monopole, Eqs. (\ref{eq:psitilde}) - (\ref{eq:stilde}), 
into the above expressions, we obtain the charge densities for the spin 1/2 monopole.
Using Eqs. (\ref{eq:higgs}) and (\ref{eq:gauge}), we find
\begin{align}
\rho_{A1} &= -\frac{2}{eM}K(H+K^2-1)r^{-4}\hat{r}^kJ^k, \\
\rho_{A2} &= -\frac{2}{eM}(-K+K^2)(H+K^2-1)r^{-4}\hat{r}^kJ^k, \\
\rho_\psi  &= -\frac{4}{eM}H^2K^2r^{-4}\hat{r}^kJ^k ,\\
\rho_S &= \rho_P=0,
\end{align}
where $H$ and $K$ are given by Eqs. (\ref{eq:H}) and (\ref{eq:K}), respectively,  with $y=evr$.

Now, it is easy to calculate the dipole moment of the electric charge distribution.
For example, the contribution of the fermion can be written as
\begin{align}
d_\psi^i &= \displaystyle\int dV r^i\rho_\psi      \notag \\
         &= \displaystyle -\frac{2}{eM}\frac{4\pi }{3}\int dr 2r^{-1}H^2K^2 J^i.
\end{align}
Then, noting that $eQ_m=4\pi $, we can calculate the $g$-factor due to the fermion as
\begin{align}
g_\psi &= \displaystyle\frac{4}{3 }\int^\infty_0 dr 2r^{-1}H^2K^2                                        \notag\\
       &= \displaystyle\frac{8}{3}\int^\infty_0 dy \frac{y (y \cosh y -\sinh y)^2}{\sinh^4 y}           \notag \\
       &= \displaystyle \frac{4}{3}\zeta(3) - \frac{4}{9}                                                 \notag\\ 
       &= \displaystyle1.158298093\cdots.                                                         \label{eq:gpsi}
\end{align}
We note that the $\zeta$-function appears in the above expression. It is interesting that the share of 
the fermion contribution to the $g$-factor is not given by a simple fractional number.

We can calculate the $g$-factors corresponding to $\rho_{A1} $ and $\rho_{A2}$ similarly.
Doing so, we obtain
\begin{align}
g_{A1} &=\displaystyle \frac{4}{3}\int^\infty_0 dy  \frac{(y^2+y \cosh y\sinh y -2\sinh^2 y)}{\sinh^3 y }    \notag \\
       &=\displaystyle -\frac{7}{3}\zeta(3) + \frac{10}{3}                                                    \notag\\
       &=\displaystyle 0.5285338925\cdots,                                                           \label{eq:ga1}  \\
\intertext{and}
g_{A2} &=\displaystyle \frac{4}{3}\int^\infty_0 dy \frac{ (y-\sinh y)(y^2+y \cosh y\sinh y -2\sinh^2 y)}{\sinh^4 y } \notag \\
       &=\displaystyle \zeta(3) - \frac{14}{9}                                                                      \notag\\ 
       &=\displaystyle -0.3534986523\cdots.                                                           \label{eq:ga2} 
\end{align}

From Eqs. (\ref{eq:gpsi})-(\ref{eq:ga2}), we find that the sum of the three contributions to the $g$-factor is given by
\begin{equation}
g_\rho =g_\psi +g_{A1}+g_{A2}=\frac{4}{3}.
\end{equation}
This means that the dipole moment of the electric charge distribution is only 2/3 of the dipole moment
determined by the long-distance behavior of the electric field.

The fact that the coefficient of the dipole field and the dipole moment of the charge distribution differ 
is not necessarily a surprise.
For the Gauss-law relation, $\vec{\nabla}\cdot \vec{ E} =\rho$, we have
\begin{align}
\int dV \ \vec{r}\ \rho &=\int dV \ \vec{r}\ \vec{\nabla}\cdot \vec{ E}                               \notag\\
                        &=\lim_{r \to \infty} \oint d\Omega \vec{r}r^2E_r-\int dV \vec{E}            \notag\\
                        &=\frac{2}{3}\vec{d}_\infty -\int dV \vec{E} .                        \label{eq:gauss}
\end{align}
In the last equality, we have assumed the following long-distance behavior of the electric field:
\begin{equation}
E^i\rightarrow
\frac{(3\hat{r}^i\hat{r}^j-\delta^{ij})}{4\pi r^3}\ d^j_\infty.
\end{equation}
This implies that the dipole moment of the charge distribution is 2/3 of the dipole moment determined by
the long-distance behavior if the volume integral of $E^i$ vanishes.
Actually, this is the case for the present system. 
Because the electric field is proportional to $ (3\hat{r}^i\hat{r}^j-\delta^{ij})J^j$ in the entire space,
the last term of Eq. (\ref{eq:gauss}) vanishes.
In the next section, we discuss the origin of the remaining 1/3 of the dipole moment.

\section{The contribution of magnetic current}

In this section we calculate the contribution of the magnetic current to the electric dipole moment.
Obviously, a loop of magnetic current will generate an electric dipole moment.
In order to extract the expression for the magnetic current, we start with the  Bianchi identity, 
\begin{equation}
D_0B^i+\epsilon^{ijk}D_jE^k=0 .
\end{equation}
For a static field configuration, this can be written, after projection onto the direction of $\hat{S}^a$, as  
\begin{equation}
\epsilon^{ijk}\partial_j(\hat{S}^aE^{ak} )=-j_1^i-j_2^i-j_3^i, \label{eq:current}
\end{equation}
where
\begin{align}
j_1^i&=-\epsilon^{ijk}(\partial_j \hat{S}^a)E^{ak},        \\ 
j_2^i&=e\epsilon^{ijk}\epsilon^{abc}\hat{S}^aA_j^bE^{ck},  \\   
j_3^i&=e\epsilon^{abc}\hat{S}^aA_0^bB^{ci} . 
\end{align}
The right-hand side of Eq. (\ref{eq:current}) can be regarded as the magnetic current, up to the sign.
The minus sign follows from a generic property of the duality transformation.
Decomposition into the $j_i$ is carried out according to the origin of the definition of each piece.
Note that $j_1^i+j_2^i$ is gauge invariant, and $j_3^i$ is invariant under time independent gauge
transformations.

Substituting the field configuration for the spin 1/2 monopole, we have
\begin{align}
j_1^i&=\frac{1}{eM}K(H+H^2+K^2-1)r^{-4}\epsilon^{ijk}J^k\hat{r}^k,         \\ 
j_2^i&=\frac{1}{eM}(-K+K^2)(H+H^2+K^2-1)r^{-4}\epsilon^{ijk}J^k\hat{r}^k,  \\   
j_3^i&=\frac{1}{eM}H^2K^2r^{-4}\epsilon^{ijk}J^k\hat{r}^k .   
\end{align}
Because the electric dipole moment generated by the magnetic current can be expressed as  
\begin{equation}
d^i=-\frac{1}{2}\int dV\epsilon^{ijk}r^jj_m^k ,
\end{equation}
we have the following results for the contribution of each magnetic current to the $g$-factor
\begin{align}
g_{j1}&=\displaystyle\frac{2}{3}\int_0^\infty dy \frac{y^2-\sinh^2y-y\cosh y\sinh y+y^2\cosh^2y}{\sinh^3 y}         \notag \\
      &=\displaystyle\frac{4}{3},                                                                                   \label{1} \\ 
g_{j2}&=\displaystyle\frac{2}{3}\int_0^\infty\!\!\! dy 
                     \frac{(y\!-\!\sinh y)(y^2\!-\!\sinh^2y\!-\!y\cosh y\sinh y\!+\!y^2\cosh^2y)}{\sinh^4 y}          \notag \\
      &=\displaystyle -\frac{1}{3} \zeta(3) - \frac{5}{9}                                                              \notag\\ 
      &=-0.956241190\cdots,                                                                                       \label{2}  \\
g_{j3}&=\displaystyle\frac{2}{3}\int_0^\infty dy \frac{y(\sinh y-y\cosh y)^2}{\sinh^4 y}                              \notag \\
      &= \displaystyle\frac{1}{3}\zeta(3) -\frac{1}{9}                                                                \notag \\ 
      &=0.2895745233\cdots.                                                                                         \label{3}         
\end{align}
We note that the sum of these three contributions is given by
\begin{equation}
g_{j1}+g_{j2}+g_{j3}=\frac{2}{3},
\end{equation}
which is what we expected.

\section{Conclusion}

We have seen that in the $N=2$ supersymmetric gauge theory, the electric dipole moment
of the spin 1/2 magnetic monopole takes the value inferred from the Dirac equation.
Although the arguments given here are based on the classical solution of the equations of motion, 
the results would not be changed by radiative corrections because of the 
supersymmetry of the system. 
In connection to this point, it would be interesting to determine the effects of supersymmetry breaking.
This may change the results through the radiative corrections, 
as well as through the modification of the classical solutions. 
These points are under investigation.

We found that the source of the electric dipole moment consists of various 
components.
In particular, the fraction of the contribution of the fermion fields is given by
a curious number.
This fraction should be independent of the gauge choice, because $\rho_{\psi}$
is gauge invariant.

We also found that 1/3 of the electric dipole moment is generated by the magnetic
current.
As we have seen, this is a rather generic result for soliton solutions that satisfy
a certain symmetric property.
Therefore, the fraction of the magnetic current contribution would be the same
even if the value of the electric dipole moment deviated from the Dirac moment,
for example, in non-supersymmetric models.

The structure of the source of the electric dipole moment which we have discussed in the
present paper would be obscured in the dual description of magnetic monopoles, if such exists,
because in such a description, magnetic monopoles are treated as local fields.
This, in turn, implies that even the magnetic moment of an ordinary electrically charged
particle may have similar structures in its source.
It is an interesting question whether this has any observable effects.

\section*{Acknowledgement}

The author would like to thank N. Ishibashi, T. Kugo, I. Kishimoto and Y. Okada for valuable discussions.

%


\end{document}